# Detection of ultra-low concentration $NO_2$ in complex environment using epitaxial graphene sensors

*Christos Melios[1,2]\*, Vishal Panchal[1], Kieran Edmonds[3], Arseniy Lartsev[4], Rositsa Yakimova[5], and Olga Kazakova[1]*

[1]National Physical Laboratory, Teddington, TW11 0LW, UK
[2]Advanced Institute of Technology, University of Surrey, Guildford, GU2 7XH, UK
[3]Royal Holloway, University of London, Egham, TW20 0EX, United Kingdom
[4]Formerly: Chalmers University of Technology, Gothenburg, S-412 96, Sweden
[5]Linköping University, Linköping, S-581 83, Sweden

\*christos.melios@npl.co.uk

**Abstract**

We demonstrate proof-of-concept graphene sensors for environmental monitoring of ultra-low concentration $NO_2$ in complex environments. Robust detection in a wide range of $NO_2$ concentrations, 10-154 ppb, was achieved, highlighting the great potential for graphene-based $NO_2$ sensors, with applications in environmental pollution monitoring, portable monitors, automotive and mobile sensors for a global real-time monitoring network. The measurements were performed in a complex environment, combining $NO_2$/synthetic air/water vapour, traces of other contaminants and variable temperature in an attempt to fully replicate the environmental conditions of a working sensor. It is shown that the performance of the graphene-based sensor can be affected by co-adsorption of $NO_2$ and water on the surface at low temperatures (≤70 °C). However, the sensitivity to $NO_2$ increases significantly when the sensor operates at 150 °C and the cross-selectivity to water, sulphur dioxide and carbon monoxide is minimized. Additionally, it is demonstrated that single-layer graphene exhibits two times higher carrier concentration response upon exposure to $NO_2$ than bilayer graphene.

Nitrogen dioxide ($NO_2$) is a chemical compound released into the atmosphere as a pollutant when fuels are burned in petrol and diesel engines. Several studies have shown that $NO_2$ can be harmful to people when inhaled for a prolonged period, resulting in airway inflammation[1–3]. In response, both the European Union (EU First Daughter Directive (99/30/EC)[4] and the UK's Department for Environment, Food and Rural Affairs (Air Quality Strategy (2000))[1] established legislation standards, in an attempt to minimise the prolonged effects of $NO_2$ inhalation[4]. In this legislation, the European Commission suggests an hourly and an averaged annual exposure to $NO_2$ concentration of 200 µg/m³ (~106 parts per billion (ppb), not to be exceeded 18 times per year) and 40 µg/m³ (~21 ppb), respectively. However, in central London for example, the monthly average $NO_2$ concentration for 2017, ranges from 34.2 to 44.1 ppb (figure 1)[5], much higher than the legislated standard limit. This signifies an urgent need for a high sensitivity, low cost and low energy consumption miniaturised gas sensor to carefully monitor the $NO_2$ levels in a broadly distributed sensor network, which will help enforce regulations. Currently optical techniques such as chemiluminescence are used for environmental monitoring, however their high capital and operating costs are a limiting factor[6]. Metal-oxides are also currently employed as a sensing material in low cost sensors. However, they operate typically in the ppm regime and suffer from high energy consumption[6–8]. One exception is the sensors described in Ref.[9], which were modified to improve the signal-to-noise ratio (and which utilise membranes to improve selectivity). Other sensing nanomaterials involve polyaniline (PANI) nanocomposites[10], carbon nanotube thin films[11] and silicon nanowires[11], however, the sensitivity and performance of sensors made of these nanomaterials depends highly on the material preparation. Moreover, these sensors demonstrate sufficiently high sensitivity only in ppm regime. Graphene has already demonstrated great potential in gas sensing, particularly for $NO_2$ molecules[12–20], therefore successful implementation of a graphene-based sensor can provide straightforward environmental pollution monitoring, miniaturised detectors suitable for portable operation and even wearable, automotive and mobile sensors for a global real-time monitoring network.

Various experimental and theoretical studies have shown that the electrical conductivity of graphene is sensitive to adsorption of gas molecules down to ppb level [21–23] and even single $NO_2$ molecule detection has been demonstrated[24]. This exceptional sensitivity is attributed to the high adsorption ability and surface-to-volume ratio of graphene, which makes graphene an ideal material for gas sensing applications. Although these extreme sensitivities are highly desirable for a gas sensor, the change in electronic properties from natural variations of ambient humidity can greatly affect the operation of devices in the ambient air[25]. Nevertheless, studies of the specific gas sensitivity at the low 10 ppb range were rarely performed in a complex environment, which would mimic the real outdoor/indoor conditions[26,27]. In practice, an integrated graphene-CMOS $NO_2$ sensor was recently demonstrated, however, the sensitivity in the ppm regime makes it unsuitable for environmental monitoring[28].

A promising method for graphene growth is via thermal decomposition of SiC[29–31]. This method is capable of producing large-area graphene directly on semi-insulating SiC substrate, which is ideal for electronic integration, eliminating the need for post-growth transfer. In this type of graphene, the interfacial layer (IFL), which is a layer of sp$^2$ and sp$^3$ bonded carbon atoms, provides strong electron doping, which can reach ~$10^{13}$ cm$^{-2}$ (in the pristine state in vacuum)[32,33]. However, the electron concentration decreases to ~$10^{12}$ cm$^{-2}$ when the sample is left in ambient air for a prolonged period of time, i.e., several days.[23] Approximately half of the reduction in the electron carrier density was previously attributed to p-doping, e.g. from water vapour and NO$_2$ present in the atmosphere, with different sensitivities among 1 and 2LG[25,34–37].

Although several works reported the effects of doping of graphene due to the presence of NO$_2$[24,27,38–40], there are currently no comprehensive studies demonstrating the combined impact of NO$_2$ and water on the electronic properties of 1LG and 2LG as well as the changes in the sensor performance due to temperature fluctuations. In this work, we systematically investigate the changes in electronic properties of 1LG and 2LG Hall crosses upon exposure to synthetic air (SA), i.e., a mixture of O$_2$ (21.28%) balanced with N$_2$, water vapour and NO$_2$ at concentrations similar or lower than those occurring in ambient air and the cross-selectivity to SO$_2$ and CO (other contaminants present in the ambient air). We perform our measurements by precisely controlling the environment that the graphene-based sensor is exposed to: from vacuum ($10^{-7}$ mbar) to NO$_2$ concentrations ranging from 10-154 ppb (i.e. the typical range required for environmental monitoring) at various temperatures as well as a combination of water vapour, SA and NO$_2$ in an attempt to replicate fluctuations in the working environment. In these experiments, we simultaneously measure the carrier concentration of both 1 and 2LG as well as 4-terminal resistance and carrier mean free path, an important electrical property providing essential information about doping and impurity scattering at the different NO$_2$ concentrations. The results reveal ultra-high response of graphene devices, down to 10 ppb NO$_2$, even in complex environmental conditions at a wide temperature range, combined with great repeatability, demonstrating the potential of graphene-based devices in NO$_2$ sensing.

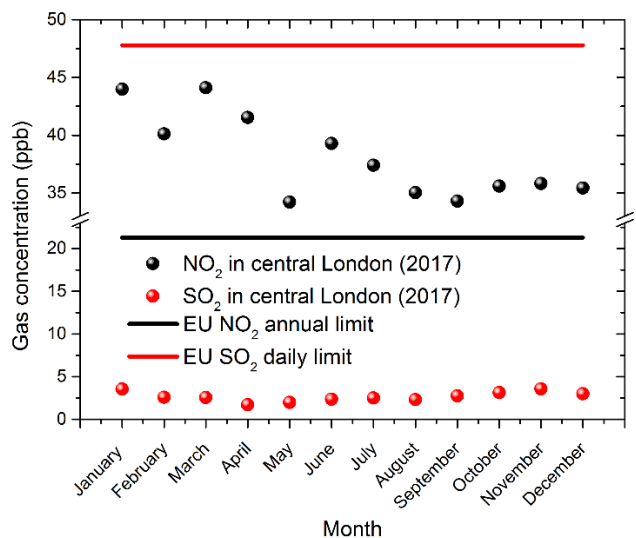

*Figure 1: (a) Monthly average of $NO_2$ (black dots) and $SO_2$ (red dots) concentration levels for 2017 in central London, UK. The black lines indicates the EC annual limit for exposure to $NO_2$ and the red line indicates the daily average limit of exposure to $SO_2$ (not to by exceeded 3 days per year)[5].*

## Methods

### Sample preparation

Epitaxial graphene on SiC was grown on semi-insulating 6*H*-SiC(0001) commercial substrates (II-VI, Inc.) with resistivity >1010 Ω cm$^{-1}$. The substrates were 8×8 mm$^2$ and misoriented ~0.05° from the basal plane. Graphene was synthesised via Si sublimation from SiC using an overpressure of Ar inert gas. Prior to the growth, the substrate was etched in $H_2$ at 100 mbar using a ramp from room temperature to 1580 °C to remove polishing damage. At the end of the ramp, the $H_2$ was evacuated, and Ar added (the transition takes about 2 minutes). Graphene was then synthesised at 1580 °C for 25 min in an Ar atmosphere. Afterwards, the sample was cooled in Ar to 800 °C.

The device was fabricated using a three-step process. Step 1: the electrical contact pads were defined using electron beam lithography (EBL), oxygen plasma ashing and electron beam physical vapour deposition (EBPVD) of Cr/Au (5/100 nm). This ensured robust contact to the SiC substrate. Step 2: the electrical leads were defined using EBL and EVPVD of Cr/Au (5/100 nm). This ensured good electrical contact to graphene. Step 3: the Hall bar design was defined using EBL and oxygen plasma etching. To ensure pristine graphene surface following the sensor fabrication, residual Poly(methyl methacrylate) was removed using contact mode atomic force microscopy. The width and length (cross-to-cross) of the device are 1 and 2.8 μm, respectively.

## Magneto-transport measurements

The global transport properties of the 1LG and 2LG Hall bar device were determined by measuring the carrier density and mobility using the AC Hall effect and 4-terminal resistance (Figure 2b). The AC Hall effect was induced by a coil that produced an AC magnetic field ($B_{AC}$ = 5 mT) at a frequency of $f_{coil}$ = 126 Hz. The resulting Hall voltage ($V_H$) response of the DC current biased ($I_{bias}$ = 50 μA) device was measured using lock-in amplifiers referenced to the first harmonic of $f_{coil}$. The electron carrier density was defined as $n_e = I_{bias}B_{AC}/eV_H$, where $e$ is the electron charge. The channel resistance ($R_{ch}$) was determined by using the 4-terminal technique, $R4 = (V_1-V_2)/I_{bias}$, where $V_1-V_2$ is the voltage drop from cross 1 to cross 2, measured using a digital voltmeter. The 4-terminal technique excludes the contact resistance, thus enabling accurate measurement of the graphene channel with well-defined length ($L$) and width ($W$). The carrier mean free path was calculated using $\lambda = (h\mu/2e)\sqrt{n/\pi}$, where $h$ is Plank's constant and $\mu = (L/W) \times (1/R4en)$. See Ref. [23] for further details on the global transport measurement techniques.

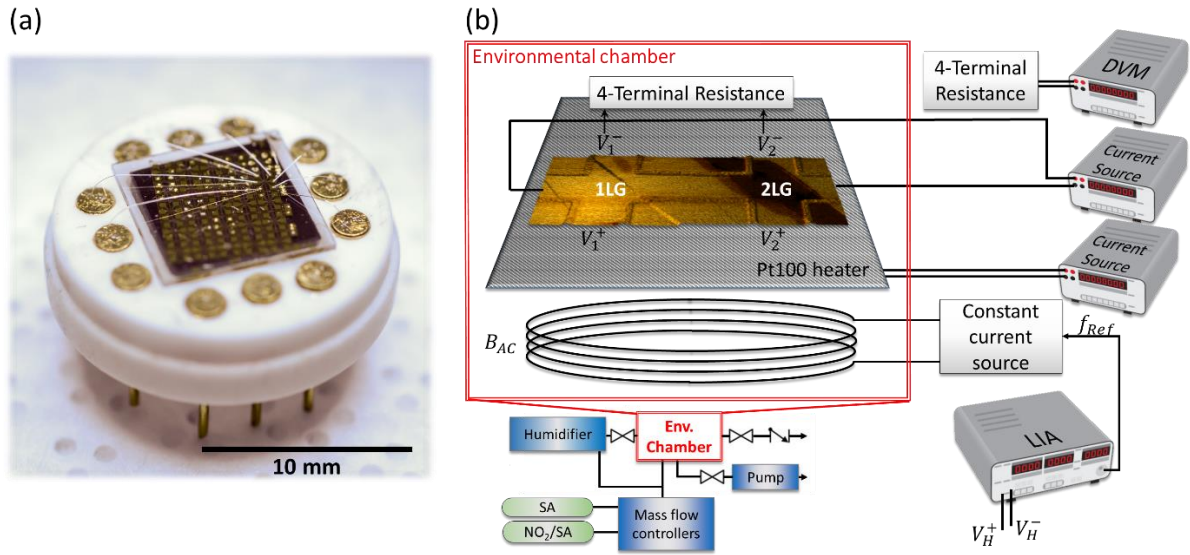

*Figure 2: (a) Picture of fabricated epitaxial graphene chip featuring 25 sensor devices on a ceramic TO-8 header with a Pt-100 heater attached. (b) Schematic of the experimental set-up for measurements of transport characteristics in the environmental chamber using a lock-in amplifier (LIA), digital voltmeter (DVM) and current source. The red box shows the environmental enclosure.*

## Environmental control

The graphene device was mounted on a ceramic TO-8 header attached to a platinum thin film heater (Pt-100), controlled by a PID feedback loop, allowing precise temperature control (70-200 °C). For the magneto-transport measurements, an in-house environmental transport measurement system was developed, equipped with two mass flow controllers (MFC), a humidifier, and a turbo-molecular vacuum pump allowing pressures of P≈$10^{-7}$ mbar. The first MFC was connected to a SA cylinder, containing $N_2$, balanced with 21.28% $O_2$ and <1ppm $CO_2$ (this value is insignificant compared to the hundreds of ppm measured in ambient air), whereas the second MFC was connected to a high purity 262 ppb $NO_2$ cylinder, balanced with SA (the gas concentration in the cylinder was certified by BOC Limited using standards traceable to ISO standards with uncertainty of ≤5%). The dilution of the $NO_2$ gas was achieved by carefully controlling the flow rates of the two gases, while maintaining a total flow rate of 1L/min into the chamber. Before each $NO_2$ exposure, the sample was annealed at 170°C overnight to ensure the clean state of the device.

# Results and discussion

## Dry $NO_2$/Synthetic air

A surface potential map of the device is displayed in figure 3(f), showing the two Hall crosses (cross 1-1LG and cross 2-2LG). In addition, a small inclusion of 3LG (darker contrast) is shown, however it does not contribute to the measurements as it is outside of the sensing area. Prior to each measurement, the sample was annealed at 170 °C in vacuum ($P\sim10^{-7}$ mbar) overnight and allowed to cool down to 70 °C. The vacuum annealing process is vital for removing any adsorbed molecules (i.e., $H_2O$, $O_2$ and $NO_2$ remaining from former runs), therefore negating any previous environmental doping effects. The magneto-transport properties in this pristine state at 70 °C (referred to as the control) are the following: carrier density of 1LG and 2LG were $n_e^{1LG}$ = 9.7×$10^{12}$ cm$^{-2}$ and $n_e^{2LG}$ = 1.2×$10^{13}$ cm$^{-2}$, respectively; the resistance of the 1-2LG channel was $R4$ = 3.3 kΩ, which combined with the weighted arithmetic mean carrier density (64% contribution from $n_e^{1LG}$ and 36% from $n_e^{2LG}$ by area), translates to an average channel carrier mobility and mean free path of $\mu_e$ = 683 cm$^2$/Vs and 24.8 nm, respectively.

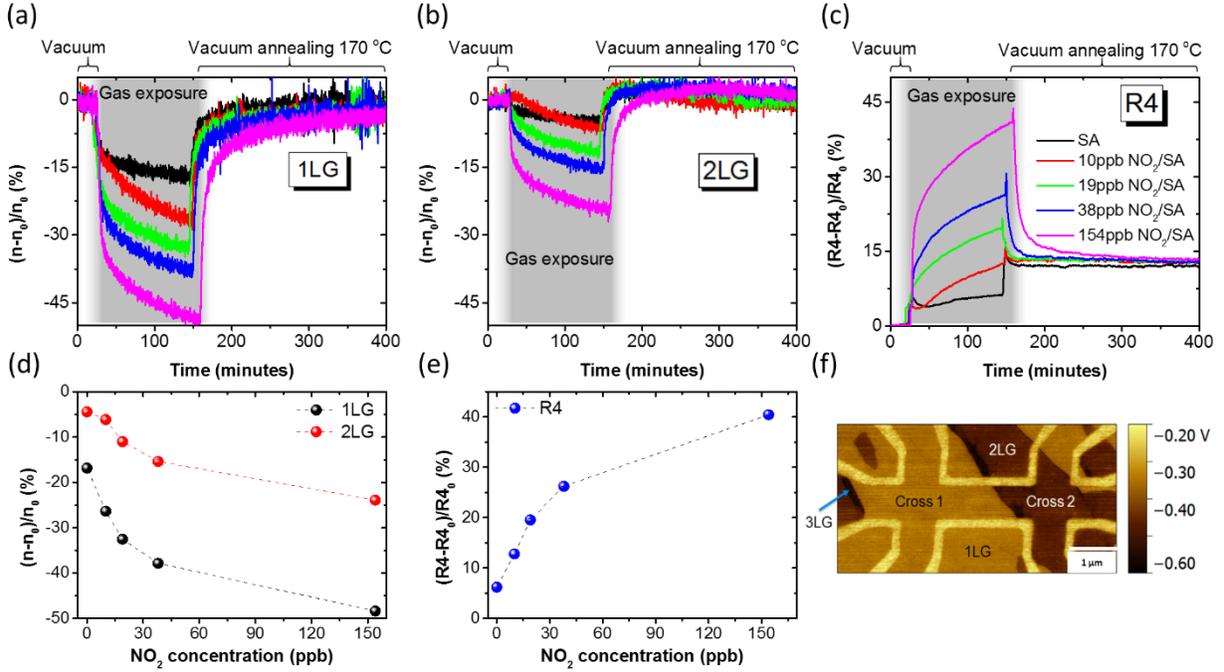

*Figure 3: (a-b) Time-dependent relative electron concentration changes for 1LG and 2LG, respectively, and (c) relative R4 changes for different $NO_2$ concentrations at 70 °C. (d) The relative changes in the carrier concentration for 1LG/2LG (black/red) and (e) R4 dependence on the $NO_2$ concentration. These average values were obtained from (d-e) after 2 hours of exposure. (f) Surface potential map of the graphene device in vacuum showing the structure of crosses 1 and 2 as 1LG and 2LG, respectively, with some presence of 3LG on the channel only.*

Synthetic air was used as carrier gas for diluting the $NO_2$ (262 ppb) to low concentrations, while mimicking dry ambient air. For consistency and practical reasons, in each measurement cycle, the graphene device was exposed to the gas mixture for ~2 hours, allowing the sensor to reach an almost steady-state. However, it is worth noting that the electronic properties of graphene will continue to marginally change, as long as the sample is exposed to the gas. Figure 3 (a and b) and table 1 show the changes in the electron concentration for 1 and 2LG, respectively, for 10-154 ppb of $NO_2$. The electron concentration of 1 and 2LG exhibits a decrease of ~16% and ~4% when the device is exposed to SA, in excellent agreement with the previous work by Panchal et al.[34]. Subsequently, the sample was annealed in vacuum (170 °C) in order to restore it to pristine condition. In figure 3c the momentary sharp spikes around the ~150 minute is due to the increase in annealing temperature, which causes the carrier mobility to decrease and thus leading to increase in resistance. Thereafter, as the adsorbates are getting removed from the graphene surface at the high temperature, the resistance drops and eventually stabilises after ~400 minutes. Following the restoration of the sample by annealing in vacuum, higher concentrations of NO$_2$/SA mixtures were introduced into the chamber. In each exposure cycle, both the carrier concentration and resistance exhibit a fast change (decreased carrier concentration and increased resistance) in the first ~10 minutes, followed by approaching a steady-state regime. Our results are consistent with the theoretical predictions of Leenaerts *et al.*[41], which demonstrated that NO$_2$ acts as a p-dopant on graphene. However, after 2 hours of exposure, 1LG exhibits about 2 times higher response compared to 2LG (i.e. ~47% decrease in electron concentration, compared to ~ 23% for 2LG, when the device was exposed to the highest NO$_2$ concentration of 154 ppb). Since the graphene-molecule interactions are strongly governed by the graphene-substrate interactions[25], it is expected that in the case of *AB*-stacked 2LG (in which case the additional graphene layers screen the substrate interactions more effectively than 1LG) the graphene-molecule electrostatic interactions will be less pronounced[34,40]. However, we cannot exclude the effects of difference in band structure between 1LG and 2LG[37], or the existence of a small band gap in the case of 2LG[42,43]. The summarised relative changes in carrier concentration for 1LG, 2LG and R4 for the different NO$_2$ concentrations are plotted in figure 3 (d and e). Both carrier concentration and R4 plots demonstrate a monotonic response, similarly to Ref. [13], with detection limit below 10 ppb. Although Density functional theory (DFT) simulations are required for more conclusive description, we can propose the following possible mechanism which contributes to the non-linear response: The NO$_2$ adsorption takes place at different adsorption sites (*i.e.* low-binding energy adsorption sites, such as sp$^2$ bonded carbon, and high-binding energy adsorption sites, such as defects). At low NO$_2$ concentrations, the high energy adsorption sites get occupied first, while at higher NO$_2$ concentrations the low energy sites start to contribute[27]. Therefore, a "competitive" mechanism between molecules takes place. However, saturation of the device cannot be rulled-out, despite previous experiments of epitaxial graphene that demonstrated response to even ppm concentrations of NO$_2$[44,45]. In addition to the carrier concentration and R4 measurements, the mean free path of the charge carrier was calculated for the different gas exposures (Table 1). The electrons in pristine graphene (vacuum at 70 °C) travel ~25 nm before scattering. However, increase in the charge carrier scattering from NO$_2$ molecules (at 154 ppb) decreased this further to 19 nm.

Since future graphene-based gas sensors will operate in a real environment (*i.e.* with variations in temperature), the doping effects of NO$_2$ on epitaxial graphene were investigated in temperature-dependent measurements. Similarly to the previous measurements, before each gas exposure, the graphene device was annealed in vacuum at 170 °C and then left to reach equilibrium at set temperatures (70, 100 and 150 °C). Unless stated otherwise, all of the measurement changes (i.e. sensor response) are relative to the specific pristine state at each temperature. The summary of the temperature-dependent measurements is shown in figure 4. These contour plots are summarised by extracting four examples along the dashed lines (indicated as i, ii, iii, iv) with the values shown in Table 1. Point (i): when the graphene device is at 70 °C and exposed to 38 ppb NO$_2$ for 2 hours, the carrier concentration of 1 and 2LG exhibits a decrease by ~37% and ~14%, respectively, while the R4 of the graphene device increased by 25%, compared to the control state. Moving along the red dashed lines of figure 4 towards point (ii), the sensor was exposed to the same $NO_2$ concentration, but at higher temperature (150 °C). The device exhibited higher response, in which case the electron concentration of 1 and 2LG decreased by a total of ~47% and ~31%, respectively, when compared to the control state. It signifies that at high temperatures $NO_2$ adsorption occurs more efficiently (however, at a higher critical temperature, the desorption rate will be higher than the adsorption). Similar results were previously demonstrated in Refs. [12,45]. At 150 °C, the mean free path decreased to 16 nm, compared to ~21 nm at 70 °C, due to higher phonon scattering at increased temperature. At point (iii) of the contour plots in figure 4, the device was exposed to the highest concentration of $NO_2$ (154 ppb), while the temperature was kept at 70 °C. In this case, the electron concentrations of 1 and 2LG demonstrated a total decrease of ~47% and ~23%, respectively, when compared to the control state. At this point, increase in impurity scattering due to the presence of $NO_2$ molecules in the graphene surface further decreased the mean free path of the electrons to 19 nm. The last point (iv) represents the highest $NO_2$ concentration (154 ppb) and highest temperature (150 °C), in which case, the electron concentration of both 1 and 2LG exhibited the highest decrease of ~54% and ~36%, respectively, compared to the control state, while the resistance of the device demonstrated the largest increase by ~76%. At this stage, there are three dominant scattering mechanisms: (i) electron-electron interactions in the graphene layer, (ii) electron-lattice phonon scattering (present at all temperatures, but increasing with temperature) and (iii) electron-impurity scattering due to the adsorbed $NO_2$ molecules. The decrease in electron concentration (due to electron withdrawal by $NO_2$ molecules) results in lowering the electron-electron interactions. However, this mechanism is overshadowed by the increase in both impurity (from $NO_2$ molecules) and phonon scattering (higher substrate temperature), which has resulted in the lowest mean free path of ~14 nm.

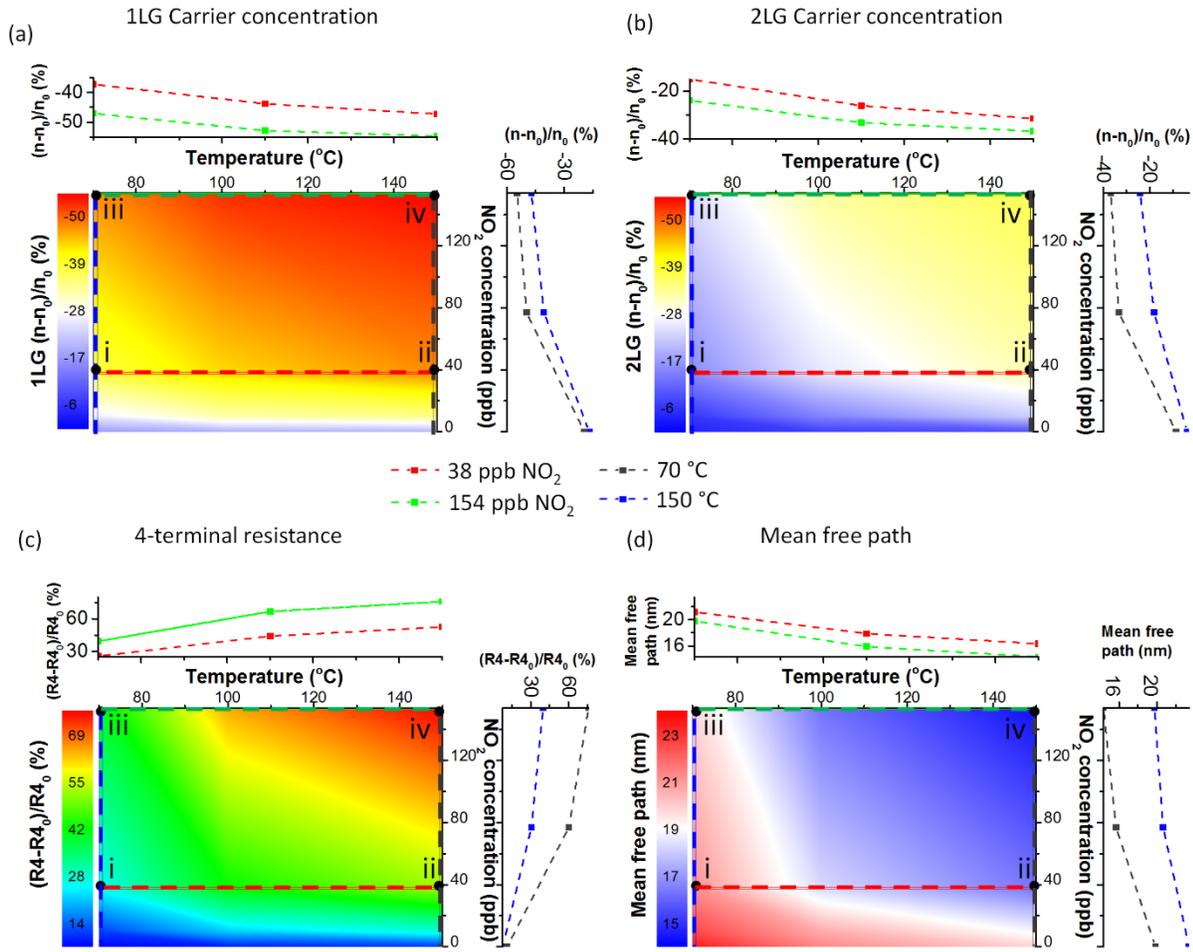

*Figure 4: Temperature-NO$_2$ concentration contour plots of (a-b) relative change in electron concentration for 1LG, 2LG, respectively; c) relative change of R4 and d) absolute values of electron mean free path. All values in (a-c) are plotted with respect to the control state of the device. Points i-iv indicate the four different examples as described in the text. The top and right insets in each panel show the resistance changes as a function of temperature (red line - 38 ppb NO$_2$; green line - 154 ppb NO$_2$) and as a function of NO$_2$ concentration (blue line - 70 °C; grey line - 150 °C)."*

*Table 1: Relative changes in electron concentration for 1LG and 2LG and R4 compared to the control state in vacuum and absolute carrier mean free path, following exposure at different $NO_2$ concentrations and temperatures.*

| Temperature (°C) | Gas | $\Delta n^{1LG}/n_0^{1LG}$ (%) | $\Delta n^{2LG}/n_0^{2LG}$ (%) | $\Delta R4/R4_0$ (%) | Mean free path (nm) |
|---|---|---|---|---|---|
| 70 | Synthetic air | -16.8 | -3.9 | 7.7 | 23 |
| 70 | 10 ppb $NO_2$/SA | -25.7 | -5.9 | 11.9 | 22 |
| 70 | 38 ppb $NO_2$/SA | -37.2 | -14.9 | 25.6 | 21 |
| 70 | 154 ppb $NO_2$/SA | -47.1 | -23.9 | 39.4 | 19 |
| 100 | Synthetic air | -18.8 | -6.7 | 9.4 | 21 |
| 100 | 10 ppb $NO_2$/SA | -31.6 | -13.9 | 21.7 | 20 |
| 100 | 38 ppb $NO_2$/SA | -43.0 | -24.7 | 42.0 | 18 |
| 100 | 154 ppb $NO_2$/SA | -52.3 | -32.3 | 64.6 | 16 |
| 150 | Synthetic air | -19.4 | -8.6 | 11.2 | 19 |
| 150 | 10 ppb $NO_2$/SA | -33.7 | -22.2 | 27.4 | 18 |
| 150 | 38 ppb $NO_2$/SA | -47.2 | -31.5 | 52.5 | 16 |
| 150 | 154 ppb $NO_2$/SA | -54.6 | -36.8 | 76.1 | 14 |

## $NO_2$ sensing in complex environments

The next set of experiments involves exposure of the graphene-based sensor to a mixture of $NO_2$ balanced with synthetic air and humidity levels 0-70% (while the sample was kept at 70 °C) in an attempt to replicate a real-life scenario of an $NO_2$ sensor, where both $NO_2$ and humidity contribute to the doping and therefore changes in resistance. Figure 5 shows the sequence of the magneto-transport measurements upon exposure to 10 and 154 ppb $NO_2$ concentrations at various humidity levels. As before, the surface of the graphene was cleaned of previously adsorbed molecules by annealing at 170 °C. The device was then allowed to cool to 70 °C and then the desired gas concentration was introduced. Following two hours of dry gas exposure, the relative humidity in the chamber was increased gradually (in 20 minutes steps). The increase in response of the sensors is clearly visible in figure 5 (a and b), where the electron concentration response increases faster following the co-adsorption of water and $NO_2$. The relative changes of electron concentration and resistance, and absolute values of the mean free path are summarised for various combinations of $NO_2$ and humidity in figure 6. A control experiment was carried out using only synthetic air (0 ppb $NO_2$), when after 2 hours of exposure the humidity was gradually increased to ~70%.

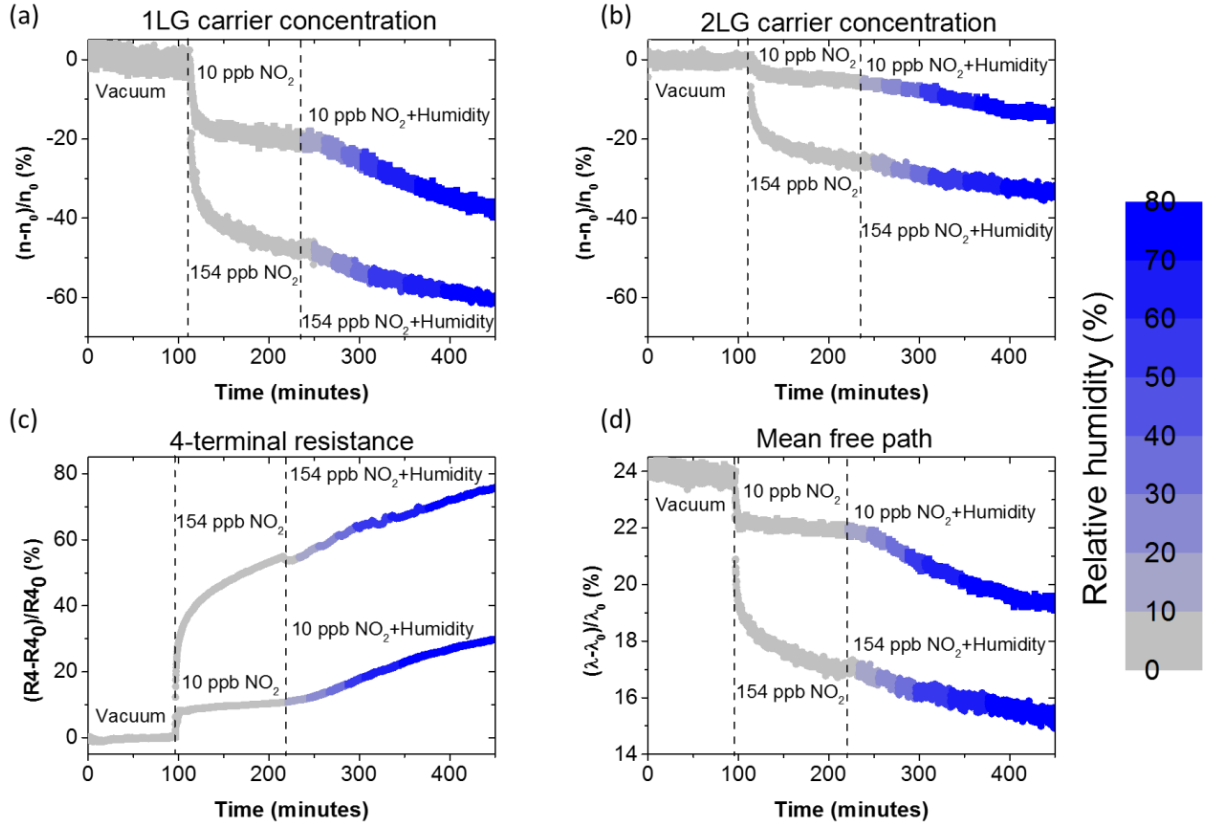

*Figure 5: Time-dependent magneto-transport measurements of (a-b) 1LG and 2LG carrier concentration, (c) 4-terminal resistance changes and (d) mean free path upon exposure to dry $NO_2$ (10 and 154 ppb) /SA (grey points) and $NO_2$ (10 and 154 ppb)/SA/humidity (light – dark blue points). The increasing intensity of the blue colour corresponds to the higher humidity, i.e. from 0 – 70%.*

First, let's consider the dry $NO_2$ case (0% R.H. in figure 6). Both electron concentrations and the resistance, demonstrate matching response to the figure 3 (d-e), highlighting the excellent repeatability of the sensor. In a dry environment, the graphene-based sensor is able to detect $NO_2$ concentrations as low as 10 ppb, with a significantly higher sensor response at 154 ppb $NO_2$ (changes along the y-axis in figure 6). Keeping the $NO_2$ concentration constant at 10 ppb, while the humidity increased to 70%, resulted in the increase in the sensor response, compared to the dry state. Although $NO_2$ molecules are strong electron acceptors on their own, which is responsible for p-doping the graphene, the combined effects of $NO_2$ and $H_2O$ molecules result in even more prominent p-doping effect. Ridene *et al.*[39] performed DFT calculations to investigate the co-adsorption of $NO_2$ and $H_2O$ on graphene. Their results demonstrated that the charge transfer from graphene to the $NO_2$ molecule is –0.10$e$ and –0.31$e$ for dry and wet exposures, respectively. This highlights the enhanced response (doping) of epitaxial graphene to $NO_2$ in wet environments. Furthermore, they suggested that during co-adsorption of $H_2O$ and $NO_2$ molecules, the lowest unoccupied molecular orbital (LUMO) of $NO_2$ is further lowered (compared to dry $NO_2$), with respect to the Fermi level of graphene. As a result, the gap between the highest occupied molecular orbital (HOMO) and LUMO of $NO_2$ is reduced[39, 46]. This process consequently leads to the enhanced charge transfer from graphene to $NO_2$. In the extreme scenario of high $NO_2$ concentrations and high humidity, the graphene device experienced the largest combined change in all three electrical properties. These measurements demonstrate that the detection of low $NO_2$ concentration ≤10 ppb can be overshadowed by the presence of high humidity (≥40% R.H.). However, this can be overcome by performing measurements at higher sensor temperatures. Figure 7 (a) shows the changes in the resistance response for the cases when the graphene is exposed to dry 10 ppb $NO_2$/SA and humid 10 ppb $NO_2$/SA for sensor temperatures 70 °C, 100 °C and 150 °C. As discussed previously, when the sensor is at 70 °C, there is a significant effect of the response due to the presence of humidity. However, the humidity effects are significantly minimized when the sensor is at higher temperatures (≥100 °C).

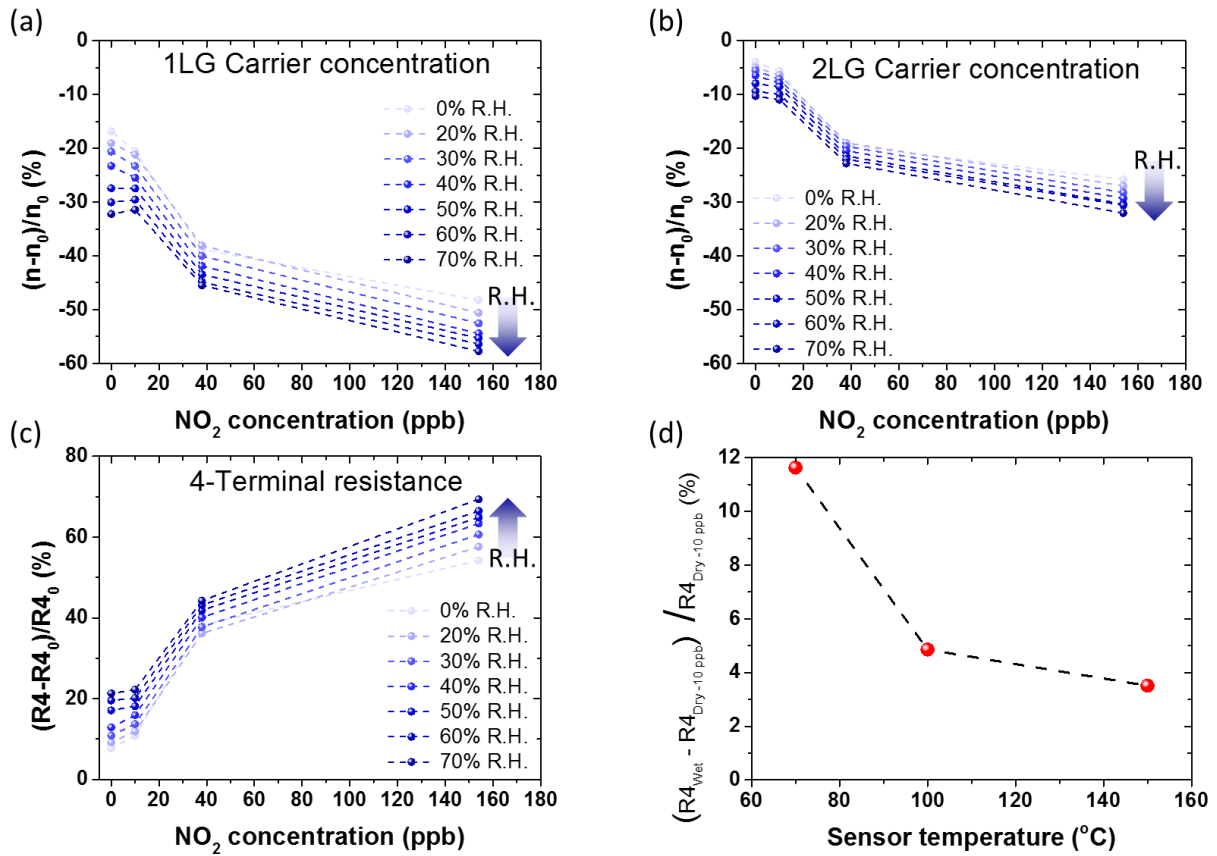

*Figure 6: Relative changes in carrier concentration for (a) 1LG, (b) 2LG and (c) R4 for the different $NO_2$ concentrations at different humidities (increase in relative humidity indicated by the arrows on the right of each plot). The averaged values were obtained 20 minutes after the*

*exposure to different humidity steps in figure 5. (d) Changes of the graphene resistance between dry 10 ppb NO₂ and 10 ppb NO₂ at 70% R.H. when the sensor is at 70 °C, 100 °C and 150 °C.*

## Sensor cross-selectivity

Since the NO₂ graphene-based sensor will operate in ambient environment, it is important to demonstrate minimum cross-selectivity between other ambient air constituents. The cross-selectivity of the graphene-based sensor to NO₂, SO₂ and CO was also studied and quantified, as shown in figure 7. At 70°C, the graphene exhibits similar sensitivities for both NO₂ and SO₂ (sensitivity to CO is considerably lower than NO₂ and SO₂), however when the sensor operates at 150°C, the sensitivity to NO₂ is dramatically enhanced to ~35 ppb/Ω (within the range of 10-38 ppb). This effectively means that at these temperatures, the sensor will be much more sensitive to NO₂ compared to SO₂. Moreover, the sensitivity to CO was found to be 5, 7 and 12 mΩ/ppb at 70°C, 100°C and 150°C, respectively. For example, if the sensor is employed in the ambient air, the resistance of the graphene will change considerably by 1.3 kΩ due to the presence of 38 ppb NO₂, but only by 57 Ω due to the presence of 2.6 ppb SO₂, 3.6 Ω for 300 ppb CO (these are typical concentrations present in the ambient air) and by 176 Ω due to the presence of 65% R.H. Lastly, CO was found to donate electrons to graphene leading to decrease in the resistance, where NO₂, SO₂ and water withdraw electrons from graphene leading to increase in resistance[41]. This demonstrates minimum cross-selectivity between the reported ambient air constituents.

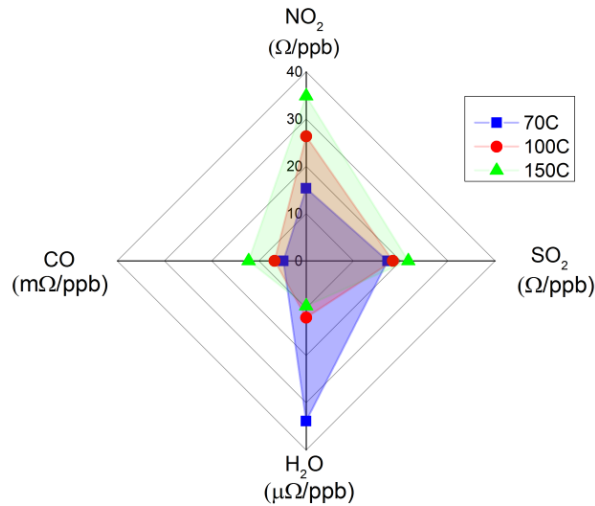

*Figure 7: Sensitivity of the graphene-based sensor for humidity, NO₂, SO₂ (at the range of 10-38 ppb) and CO (at the range of 100-1100 ppb) at 70 °C, 100 °C and 150 °C. Unit conversion note: at atmospheric pressure and air temperature ~23 °C, $R.H._{\cdot ppb} = 2.85 \times 10^5 \, R.H._{\cdot\%} - 9.55 \times 10^4$. Note the significant difference in units: NO₂ and SO₂ sensitivity is plotted in Ω/ppb, whereas CO and H₂O sensitivities are plotted in mΩ/ppb and µΩ/ppb.*

# Conclusions

In conclusion, we presented a comprehensive study of the ultra-high sensitivity of epitaxial graphene on *6H*-SiC(0001) to low concentrations of $NO_2$ (10-154 ppb, the range that is desirable for environmental monitoring). The measurements were performed in a complex gaseous environment (i.e., $N_2$, $O_2$ and humidity) and in the temperature range of 70-150 °C in an attempt to replicate a typical working environment of a graphene-based sensor. The measurements demonstrated that after being adsorbed by graphene, $NO_2$ acts as a strong electron acceptor, where 1LG donates ~2 times more electrons compared to *AB*-stacked 2LG. Consequently, 1LG is being much more sensitive to variations in the $NO_2$ concentration. This is explained by screening of the substrate-graphene interactions by the additional graphene layer.

It is also demonstrated that the response of graphene to $NO_2$ molecules can be further enhanced when the device is operated at elevated temperatures. Moreover, the combined adsorption of $H_2O$ and $NO_2$ further increased the response, allowing higher charge transfer from graphene to the $NO_2$ molecules. Lastly, it was found that the adsorption of both $H_2O$ and $NO_2$ leads to the reduction of the mean free path of electrons and therefore the increase in resistance. In our experiments, detection down to 10 ppb level of $NO_2$ was achieved. Detection of the lower $NO_2$ concentration (≤10ppb) can be masked by the presence of humidity (≥40% R.H.). However, the performance of the sensor can be improved by its operation at elevated temperatures where the effects of water are minimized, and sensitivity to $NO_2$ further improves. Moreover, at 150°C the sensor demonstrated minimum cross-selectivity to $SO_2$ and CO. These results highlight the great potential for simple-to-operate, miniaturised $NO_2$ sensors based on epitaxial graphene, with possible applications in portable devices for low-cost environmental pollution monitoring as well as automotive, mobile and wearable sensors for a global real-time monitoring network.

# Acknowledgments

The authors acknowledge the support EMPIR 2016NRM01 GRACE, NMS under the IRD Graphene Project (No. 119948) and NMS No. 121524. This project has received funding from the European Union's Horizon 2020 research and innovation programme under grant agreement GrapheneCore2 785219 number. The work was carried out as part of an Engineering Doctorate program in Micro- and NanoMaterials and Technologies, financially supported by the EPSRC under the grant EP/G037388, the University of Surrey and the National Physical Laboratory. This work was partially funded by the National Physical Laboratorys Director's Science and Engineering Fund. The authors would like to also acknowledge Alexander Tzalenchuk, Ivan Rungger for the useful discussions and Nick Martin and Matthew Berrow for the Ozone experiments.This document is the unedited Author's version of a Submitted Work that was subsequently accepted for publication in ACS Sensors, copyright © American Chemical Society after peer review. To access the final edited and published work see https://pubs.acs.org/doi/10.1021/acssensors.8b00364.# References

(1) Air Quality Expert Group. *Nitrogen Dioxide in the United Kingdom*; London, 2004.
(2) Faustini, A.; Rapp, R.; Forastiere, F. Nitrogen Dioxide and Mortality: Review and Meta-Analysis of Long-Term Studies. *Eur. Respir. J.* **2014**, *44* (3), 744–753.
(3) Mills, I. C.; Atkinson, R. W.; Kang, S.; Walton, H.; Anderson, H. R. Quantitative Systematic Review of the Associations between Short-Term Exposure to Nitrogen Dioxide and Mortality and Hospital Admissions. *BMJ Open* **2015**, *5* (5).
(4) E.C. Air Quality Standards http://ec.europa.eu/environment/air/quality/standards.htm (accessed Apr 3, 2017).
(5) King's College London. London Average Air Quality Levels https://data.london.gov.uk/dataset/london-average-air-quality-levels (accessed Jan 23, 2018).
(6) Liu, X.; Cheng, S.; Liu, H.; Hu, S.; Zhang, D.; Ning, H. A Survey on Gas Sensing Technology. *Sensors* **2012**, *12* (12), 9635–9665.
(7) Wetchakun, K.; Samerjai, T.; Tamaekong, N.; Liewhiran, C.; Siriwong, C.; Kruefu, V.; Wisitsoraat, A.; Tuantranont, A.; Phanichphant, S. Semiconducting Metal Oxides as Sensors for Environmentally Hazardous Gases. *Sensors Actuators, B Chem.* **2011**, *160* (1), 580–591.
(8) Fine, G. F.; Cavanagh, L. M.; Afonja, A.; Binions, R. Metal Oxide Semi-Conductor Gas Sensors in Environmental Monitoring. *Sensors* **2010**, *10* (6), 5469–5502.
(9) Mead, M. I.; Popoola, O. A. M.; Stewart, G. B.; Landshoff, P.; Calleja, M.; Hayes, M.; Baldovi, J. J.; McLeod, M. W.; Hodgson, T. F.; Dicks, J.; et al. The Use of Electrochemical Sensors for Monitoring Urban Air Quality in Low-Cost, High-Density Networks. *Atmos. Environ.* **2013**, *70*, 186–203.
(10) Pandey, S. Highly Sensitive and Selective Chemiresistor Gas/vapor Sensors Based on Polyaniline Nanocomposite: A Comprehensive Review. *J. Sci. Adv. Mater. Devices* **2016**, *1* (4), 431–453.
(11) Cuscunà, M.; Convertino, A.; Zampetti, E.; Macagnano, A.; Pecora, A.; Fortunato, G.; Felisari, L.; Nicotra, G.; Spinella, C.; Martelli, F. On-Chip Fabrication of Ultrasensitive NO2 Sensors Based on Silicon Nanowires. *Appl. Phys. Lett.* **2012**, *101* (10), 103101.
(12) Eriksson, J.; Puglisi, D.; Kang, Y. H.; Yakimova, R.; Lloyd Spetz, A. Adjusting the Electronic Properties and Gas Reactivity of Epitaxial Graphene by Thin Surface Metallization. *Phys. B Condens. Matter* **2014**, *439*, 105–108.16